\documentclass[twocolumn,english,aps,reprint,groupedaddress,notitlepage,nobibnotes,nofootinbib,preprintnumbers,showpacs]{revtex4-1}
\usepackage{graphicx,epsf,color,colortbl,amsmath,multirow,comment,slashed}
\usepackage[caption=false]{subfig}
\usepackage[yyyymmdd,hhmmss]{datetime}
\usepackage{enumitem}
\usepackage[bookmarks=false]{hyperref}

\newif\ifdraft
\drafttrue
\newif\ifpreprint
\preprinttrue

\def\Section#1{\noindent\textsl{#1.\/}}

\def\Fig#1{Fig.~{\ref{#1}}}

\def\Tr{\mathop{\rm Tr}}

\def\spa#1.#2{\left\langle#1\,#2\right\rangle}
\def\spb#1.#2{\left[#1\,#2\right]}

\def\eqn#1{Eq.~(\ref{#1})}

\def\Section#1{\vskip .2 cm 
\noindent{\em #1:}}

\def\spa#1.#2{\left\langle#1\,#2\right\rangle}
\def\spb#1.#2{\left[#1\,#2\right]}

\def\OS{\mathcal O_s}
\def\OL{\mathcal O_l}
\def\Oi{\mathcal O_i}
\def\Oj{\mathcal O_j}

\def\cM{\mathcal M}

\newcommand{\cG}{\cellcolor{Gray}}
\newcommand{\cR}{\cellcolor{LightGray}}
\definecolor{zero1}{rgb}{0.35,0.4,0.85}
\definecolor{zero2}{rgb}{0.88,0.88,.88}
\definecolor{zero3}{rgb}{0.85,0.85,0.95}
\definecolor{LightRed}{rgb}{1,0.8,0.8}
\definecolor{Red}{rgb}{1,0.0,0.0}
\definecolor{Gray}{gray}{0.83}
\definecolor{LightGray}{gray}{0.69}

\begin{document}

{\small \hbox{\hskip .5 cm UCLA/TEP/2019/105}  \hskip 4cm   \hbox{CERN-TH-2019-160}   }

\title{Non-renormalization and operator mixing via on-shell methods}

\author{Zvi~Bern${}^{ab}$, Julio Parra-Martinez${}^{a }$, and Eric Sawyer${}^{a }$ }
\affiliation{
  $\null$\\
${}^a$Mani L. Bhaumik Institute for Theoretical Physics,\\
UCLA Department of Physics and Astronomy,\\
Los Angeles, CA 90095, USA\\ \vspace{-0.2cm}
\\ 
 ${}^b$Theoretical Physics Department, CERN, \\                     
                         1211 Geneva 23, Switzerland 
\\ \vspace{-0.2cm}
}

\begin{abstract}
  Using on-shell methods, we present a new perturbative non-renormalization
  theorem for operator mixing in massless four-dimensional quantum field theories.
   By examining how unitarity cuts of form
  factors encode anomalous dimensions we show that longer operators are often
  restricted from renormalizing shorter operators at the first order where
  there exist Feynman diagrams. The theorem applies quite generally and
  depends only on the field content of the operators involved. We apply our
  theorem to operators of dimension five through seven in the Standard Model
  Effective Field Theory, including examples of nontrivial zeros in the
  anomalous-dimension matrix at one through four loops. The zeros at two and
  higher loops go beyond those previously explained using helicity selection
  rules. We also include explicit sample calculations at two loops.
\end{abstract}


\maketitle

\Section{Introduction} 
A key challenge in particle physics is to identify physics beyond the
Standard Model.  Because current experimental data at colliders is
well described by the Standard Model, it is unclear which theoretical
direction will ultimately prove to be the one chosen by Nature.  It is
therefore important to quantify new physics beyond the Standard Model
in a systematic, model-independent manner.  The theoretical framework
for doing so is via effective field theories that extend the Standard
Model Lagrangian by adding higher-dimension
operators~\cite{Buchmuller:1985jz,Grzadkowski:2010es}:
\begin{equation}
  \Delta \mathcal{L}= \sum_i c_i \mathcal{O}_i \,,
\end{equation}
with coefficients $c_i$ suppressed by powers of a high-energy scale $\Lambda$
dictated by the dimension of $\mathcal{O}_i$.  The resulting
theory, known as the Standard Model Effective Field Theory (SMEFT), is reviewed
in Ref.~\cite{SMEFTReview}.

As for all quantum field theories, renormalization induces mixing of these
operators. This can be parametrized by the renormalization group equation,
\begin{equation}
  16\pi^2\frac{\partial c_i}{\partial \log\mu}=\gamma^{ \rm UV}_{ij}c_j\,,
  \vspace{-0.15cm}
\end{equation}
where $\gamma^{\rm UV}_{ij}$ is the anomalous-dimension matrix and
$\mu$ is the renormalization scale. Usually, $\gamma^{\rm UV}_{ij}$ is
calculated perturbatively in the marginal couplings of the Standard
Model Lagrangian, which we will denote collectively as $g$. The
complete one-loop anomalous-dimension matrix for operators up to
dimension six has been computed in
Refs.~\cite{Grojean:2013kd,Manohar123}. These calculations reveal a
number of vanishing entries related to 
supersymmetry~\cite{ManoharHolomorphic}, which seem
surprising at first because there are valid diagrams that can be
written down.  These zeros have been elegantly
explained~\cite{NonrenormHelicity} using tree-level helicity selection
rules~\cite{ManganoReview}, which set certain classes of tree-level
amplitudes to zero. The tree-level vanishings imply through unitarity
that certain logarithms and their associated anomalous dimensions are
not present. Although these selection rules are reminiscent of
supersymmetric ones, they hold for generic massless quantum field
theories in four dimensions.  

Might it be possible that beyond one loop there are new nontrivial zeros?  
At first sight, this seems rather unlikely because the
helicity selection rules fail to hold at loop level.  In this Letter,
we show that, contrary to expectations, there are, in fact, additional
nontrivial zeros in the higher-loop anomalous-dimension matrix.  As in
Ref.~\cite{NonrenormHelicity}, our only assumption is that the theory
does not contain any relevant couplings (e.g.~masses).
To state the new nonrenormalization theorem we define the length of an
operator, $l(\mathcal{O})$, as the number of fundamental field insertions in
$\mathcal{O}$. Then the statement of theorem is as follows:
\setlist[itemize]{leftmargin=3.5mm}
\begin{itemize}[label={}]
\item
\emph{Consider operators $\OS$ and $\OL$ such that
  $l(\OL)>l(\OS)$. $\OL$ can renormalize $\OS$ at $L$ loops only if $L > l(\OL)-l(\OS)$.}
\end{itemize}
At fixed loop order, sufficiently long
operators cannot renormalize short operators because there would be
too many legs to form a diagram with the right structure. Such zeros
in the anomalous-dimension matrix are trivial.
As written above the theorem applies non-trivially at $(l(\OL)-l(\OS))$-loops, i.e., the
first loop order at which there could be renormalization because 
diagrams exist. However, in a general
theory with multiple types of fields, the first renormalization can be delayed
even further, depending on the precise field content of the two operators.
We encapsulate this into the more general rule:
\begin{itemize}[label={}]
\item
\emph{If at any given loop order, the only diagrams for a matrix
  element with the external particle content of $\OS$ but an insertion
  of $\OL$ involve scaleless bubble integrals, there is no renormalization of
  $\OS$ by $\OL$.}
\end{itemize}

What makes them nontrivial is that Feynman diagrams exist that seem as
if they should contribute to an anomalous dimension, but fail to do so
because the diagrams do not generate the appropriate logarithms. The
Feynman-diagram language can obscure this, because individual diagrams
are not gauge invariant. While not difficult to disentangle at one
loop, at higher loops it becomes more advantageous to work in an
on-shell formalism, which only takes gauge-invariant quantities as
input.  Indeed, modern unitarity methods~\cite{GeneralizeUnitarity}
have clarified the structure of loop amplitudes resulting in
significant computational advantages for a variety of problems,
including the computation of form factors and associated anomalous
dimensions~\cite{SusyAnomDims}.


\Section{Renormalization and Form Factors}
Traditionally, the anomalous dimension corresponding
to the renormalization of an operator $\Oi$ by an operator $\Oj$ is 
extracted from UV divergences. These can be found, for instance, in form factors,
\begin{equation}
  F_j[p_1,...,p_n;q;\mu]= \langle p_1,...,p_n|\Oj(q)|0\rangle\,,
\label{eq:FormFactor}
\end{equation}
with an operator insertion $\Oj$ and external states $|p_1,...,p_n
\rangle$ that overlap with states created by $\Oi$. The divergences
and associated anomalous dimensions can also be obtained from
one-particle irreducible effective actions or from scattering
amplitudes, $\mathcal{M}_{\Oj}$, corresponding to form factors
with the operator momentum injection $q$ set to zero.

Here we use the elegant on-shell approach developed by Caron-Huot and
Wilhelm that extracts anomalous dimensions directly from renormalized
quantities~\cite{Caron-Huot:2016cwu}.  In this approach, the intuition
that the renormalization properties of the theories are encoded in
on-shell form factors through their logarithms is made precise by the
following equation:
\begin{equation}
  e^{-i\pi D}F^* = S \, F^*\,,
\label{eq:ffabstract}
\end{equation}
where $F^*$ is the conjugate form factor with an insertion of 
an $\Oj$ operator.
This relates the phase of the S-matrix, $S$, to the
dilatation operator $D$ which extracts
anomalous dimensions. We point the interested reader 
to Ref.~\cite{Caron-Huot:2016cwu} for its derivation.

\begin{figure}[tb]
  \centering
\vskip -.0 cm 
  \begin{minipage}{0.45\columnwidth}
    \includegraphics[scale=0.60]{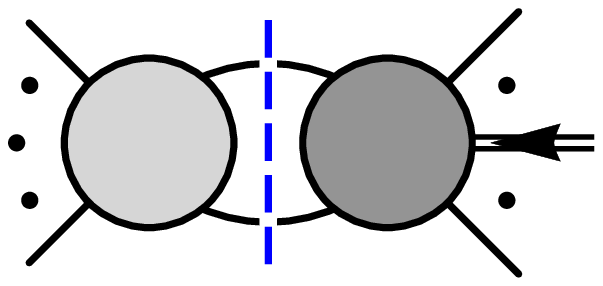}\\[-.15 cm] \hspace{-0.35cm}\textbf{(a)}\\[.02 cm]
   \includegraphics[scale=0.60]{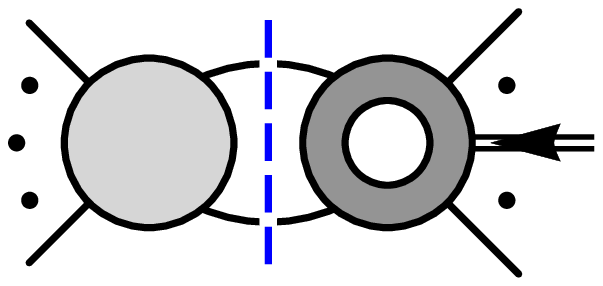}\\[-.15 cm] \hspace{-0.35cm}\textbf{(c)}
  \end{minipage}
  \begin{minipage}{0.45\columnwidth}
  \includegraphics[scale=0.60]{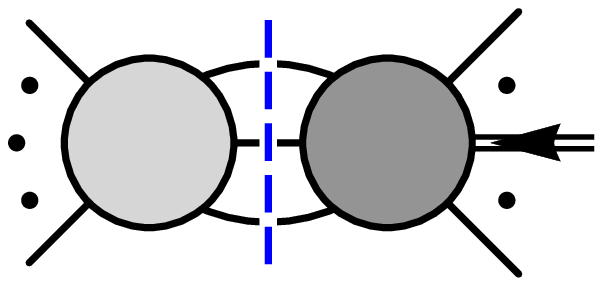}\\[-.15 cm] \hspace{-0.35cm}\textbf{(b)}\\[.02cm]
  \includegraphics[scale=0.60]{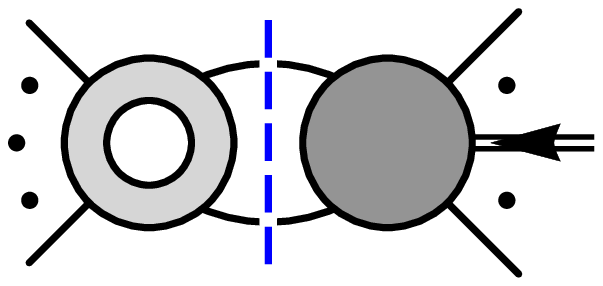}\\[-.15 cm] \hspace{-0.35cm}\textbf{(d)}
  \end{minipage}
  \caption{
    Unitarity cuts relevant for the extraction of anomalous dimensions from
    one- (a) and two-loop (b-d) form factors. The darker blobs indicate a
    higher-dimension operator insertion. The  double-lined arrow indicates the
    insertion of additional off-shell momentum from the operator.  The dashed
    line indicates the integral over phase space of the particles crossing the
    cut.
 }
  \label{fig:formfactor}
  \vspace{-0.4cm}
\end{figure}

For simplicity we use dimensional regularization.  In this case the
dilatation operator $D$ is related to the single renormalization scale
$\mu = \mu_{\rm UV} = \mu_{\rm IR}$, as $D\simeq -\mu\partial_\mu$.
Expanding Eq.~\eqref{eq:ffabstract} at one loop one obtains the
following description of the renormalization of $\Oi$ by $\Oj$
\begin{align}
\begin{split}
  & \left(\gamma_{ij}^{\rm UV}-\gamma_{ij}^{\rm IR} +\delta_{ij} \beta(g) \partial_g \right)^{(1)} \langle p_1,...,p_n|\Oi|0\rangle^{(0)}\\
 & \hspace{2.5cm}=-\frac{1}{\pi} \langle p_1,...,p_n|\mathcal M \otimes\Oj|0\rangle \,.
\end{split}
\label{eq:ffoneloop}
\end{align}
On the left hand side we find the tree-level form factor of $\Oi$, the
beta function $\beta(g)$ of the couplings $g$, the anomalous
dimensions $\gamma_{\rm UV}$, which are the objects of 
interest, and
the infrared anomalous dimensions $\gamma_{\rm IR}$, which arise from
soft and/or collinear logarithms. The superscripts denote the
perturbative order. The right hand side arises from the term $\mathcal
M F^*$, where $\mathcal M = -i(S-1)$ is the scattering amplitude.  The
notation $\otimes$ here refers to an integration over the phase space
of intermediate two-particle states in the product.  This simply
corresponds to a one-loop unitarity cut, as depicted in Figure
\ref{fig:formfactor}(a).  Schematically, \eqn{eq:ffoneloop} says that,
up to terms coming from the $\beta$ function, one-loop anomalous
dimensions are eigenvalues of the S-matrix, with the form factors being
the corresponding eigenvectors. More practically, this equation
describes how to systematically extract the anomalous dimensions from
the coefficients of logarithms by taking discontinuities of the form
factor. 

The fact that dependence on the renormalization scale is
related to discontinuities in kinematic variables is no surprise since
the arguments of logarithms must balance kinematic variables against
the renormalization scale to make them dimensionless.  This
observation has also been used to efficiently determine the
renormalization-scale dependence of the two-loop counterterm in pure
Einstein gravity from unitarity~\cite{TwoLoopGravitySimplified}.

At higher-loop orders, other unitarity cuts, matching the order of the anomalous
dimension, need to be considered. For instance at two loops, the three-particle
cut is required, as well as the two-particle cut between the tree-level
amplitude and the one-loop form factor and vice versa, as in
Fig.~\ref{fig:formfactor}(b--d).

\Section{Non-renormalization theorem}
We would like to consider the renormalization of a shorter operator $\OS$ by a
longer operator $\OL$. This could be, for example, the renormalization of
$\phi^2F^2$ by $\phi^6$, where $\phi$ is a scalar and $F$ is a vector field
strength.  For simplicity we will take $\OS$ and $\OL$ to be single operators,
though in general they represent collections of operators with the same field
content, but differing Lorentz contractions or color factors. Because our
arguments rely only on the field content and basic structure of the unitarity
cuts, our conclusions will apply just as well to the more general case.

The formalism reviewed above allows us to connect the anomalous
dimensions to unitarity cuts of form factors, given knowledge of the
$\beta$ function of the leading couplings and the infrared anomalous
dimensions. We now show that for the leading contributions,
there is an even more direct connection between the ultraviolet anomalous
dimensions and unitarity cuts.

The appearance of the $\beta$ function in Eq.~\eqref{eq:ffoneloop} is
avoided simply by extracting the anomalous dimensions from the \emph{minimal} form
factor of $\OS$, which is defined as the one with the minimum number of legs
needed to match the operator. We will denote this by a subscript on the state,
$|p_1,...,p_n\rangle_s$. Because of its defining property, the minimal
tree-level form factor is local and does not depend on the 
couplings, $g$. Therefore the dependence of the higher-loop analog of
Eq.~\eqref{eq:ffoneloop} on the $\beta$ function drops out.

Next, we would need knowledge of the infrared anomalous dimension $\gamma_{\rm
IR}$.  Infrared singularities are very well understood~\cite{IRBasicPapers,
IRLaterPapers, StermanBook, ManoharIR}.  Our case is special, with a rather
simple infrared structure.  We are interested in the first loop order at
which the higher-dimension operator could be renormalized.  This would be the
first loop order for which it is possible to write down valid diagrams. The
lack of diagrams at lower-loop order means there cannot be any $\log(\mu_{\rm IR})$ terms or
corresponding $\gamma_{\rm IR}$ at the given loop order under consideration.
In addition, infrared singularities are diagonal for the operators with
distinct fields, mixing only via color. 
Therefore at this order $\gamma_{\rm IR} = 0$.
Various examples 
will be given in Ref.~\cite{OneLoopSMEFT}.

Thus, application of \eqn{eq:ffabstract} is particularly simple 
for our case so that the relation between the first potentially nonvanishing
anomalous dimension and unitarity cuts is direct:
\begin{align}
\begin{split}
  & (\gamma_{sl}^{{\rm UV}})^{(L)}\, {}_s\langle p_1,...,p_n|\OS|0\rangle^{(0)}\\
 & \hspace{2.5cm}=-\frac{1}{\pi} {}_s\langle p_1,...,p_n|\mathcal M \otimes\OL|0\rangle \,.
\end{split}
\label{eq:ffsimplified}
\end{align}
With this relation at hand, it is now
straightforward to argue for new non-renormalization zeros by
analyzing the allowed unitarity cuts. \eqn{eq:ffsimplified} gives
$(\gamma_{sl}^{{\rm UV}})^{(L)}$ in terms of a sum over cuts of the
form illustrated in Fig.~\ref{fig:formfactor}. The left-hand side of
any such $k$-particle cut is a $n_{\cM}$-point amplitude, with the number of
particles external to the cut equal to $n_{\cM} - k$. Similarly the right-hand side is an $n_F$-point form factor,
with $n_F-k$ particles external to the cut. Now, for the minimal form
factor, the total number of external particles must match the length of
$\OS$, so we must have the relation,
\begin{equation}
n_{\cM} + n_F - 2k=l(\OS) \,.
\label{eq:legnumbers}
\end{equation}
The number of legs $n_{\cM}$ and $n_F$ are both bounded from
below. For the unitarity cut to be non-zero, the scattering amplitude
on the left must have at least two external particles, that is,
$n_{\cM}\ge k+2$. On the other side, $n_F$ is restricted by the
requirement that the form factor not include any scaleless
bubbles. Since all legs of the form factor, including those crossing
the cut, are on shell, any such scaleless bubbles would evaluate to
zero. At one loop, for example, this implies $n_F\ge l(\OL)$, which is
the same as the tree level relation. At higher loops the particle
count can be reduced depending on the number of loops in the form
factor, which produces the relation
\begin{equation}
n_F\ge l(\OL)-(L_F-1)-\delta_{L_F,0} \,.
\end{equation}
Here $L_F$ is the number of loops contained in the form
factor. $\delta_{L_F,0}$ is unity if the form factor is at tree level
and zero otherwise, which accounts for the fact that there is no
reduction in particle number between tree level and one loop. By 
considering the possible placings of the loops in the cut or on either
side of the cut, we have $L_F\le L-(k-1)$, implying $n_F\ge
l(\OL)-L+k-\delta_{L_F,0}$. Combining this with the condition on
$n_\cM$ and plugging in to equation \eqref{eq:legnumbers}, we obtain
\begin{equation}
  l(\OL)-L+2-\delta_{L_F,0}\le l(\OS)\,.
  \label{eq:lengthineq}
\end{equation}
This inequality shows that the difference in length of the operators can preclude the
renormalization unless 
\begin{equation}
  L>l(\OL)-l(\OS)\,,
\end{equation}
and thus completes the proof of the first form of our theorem.  In
summary, we have shown that at loop orders less than or equal to
$l(\OL) - l(\OS)$ there are no allowed unitarity cuts that can capture
the coefficient of $\log(\mu^2)$, which in turn implies that
$\gamma_{sl}^{\rm UV} = 0$.  Eq.~\eqref{eq:lengthineq} also shows that
the contributions to the anomalous dimension at loop order $L =
l(\OL)-l(\OS)+1$ are captured by cuts of the type in
Figs.~\ref{fig:formfactor}(a) and \ref{fig:formfactor}(b), that are
given purely in terms of tree-level matrix elements.  Cuts of the type
in Fig.~\ref{fig:formfactor}(c) are directly ruled out by
Eq.~\eqref{eq:lengthineq} and cuts of the type in
Fig.~\ref{fig:formfactor}(d) are ruled out because $l(O_l) - l(O_s) +
2$ legs need to be sewn across the cut to have a total of $l(O_s)$
external legs, so that all $l(O_l) - l(O_s) + 1$ loops are accounted
for in the cut.  This observation should help in their computation,
for instance by allowing the use of four-dimensional helicity methods to
evaluate the cut. It also implies that helicity selection rules can be active
beyond one loop, contrary to expectations. 

Depending on the particle contents of the two operators, it might happen that
there are no allowed unitarity cuts even at a higher loop order than the one
predicted by the first form of the theorem. Instead of analyzing the unitarity
cuts, this can be explained in the more familiar diagrammatic language.
Clearly, if the only diagrams that can be drawn involve scaleless bubbles,
there will be no available cut where all loops are included in the cut. Thus,
diagrams with fewer cut legs will force the form factor to include the
scaleless bubble, and thus to evaluate to zero.  Then the corresponding
anomalous dimension must also be zero.  This explains the more general rule
presented in the introduction.  As noted above this relies on the absence of
infrared singularities whenever corresponding lower-loop form factors vanish.

Examples of zeros in the SMEFT at one loop are the renormalization of $F^3$ by $\phi^2F^2$,
and of $D^2\phi^4$, $F\phi\psi^2$, and $D\phi^2\psi^2$ by $ \phi^3\psi^2$,
which were already explained using the helicity selection
rules~\cite{NonrenormHelicity}, but also follow from the principles described
here. In contrast to the helicity selection rules, however, our theorem can
also apply at higher loops.
The full set of zeros predicted by our rules for operators of
dimensions five, six and seven includes examples at one through four
loops and is described in Tables \ref{tab:dim5}, \ref{tab:dim6}, and
\ref{tab:dim7} respectively. The tables also indicate the overlap
between our theorem and the one-loop helicity selection rules of
~\cite{NonrenormHelicity}. Note we have combined some of the
categories of operators of ~\cite{NonrenormHelicity}, since our
theorem does not need to distinguish operators based on their chirality.

\begin{table}[t]
	\vspace{-0.4cm}
	\begin{center}
		\caption{Application of the non-renormalization
			theorem to dimension-five operators.  The operators
			labeling the rows are renormalized by the operators
			labeling the columns.  $\times_L$ indicates
			the theorem applies at $L$-loop order. ($L$) denotes
			that there are no diagrams before $L$-loops, but
			renormalization is possible at that order, since the
			required cuts can exist.  Light-gray
			shading indicates a zero at one loop due
			to helicity selection rules, while dark-gray shading
			indicates the entry is a new zero predicted by our
			non-renormalization theorem.}
		\vspace{0.2cm}
		\renewcommand{\arraystretch}{1.2}
		\begin{tabular}{|c|c|c|c|c|}
			\hline 
			& $F^2\phi$ & $F\psi^2$  & $\phi^2\psi^2$ & $\phi^5$ \\ 
			\hline 
			$F^2\phi$ &  &  & (2) &\cR $\times_2$  \\ 
			\hline 
			$F\psi^2$ &  &  &\cG $\times_1$ &\cR $\times_3$ \\ 
			\hline 
			$\phi^2\psi^2$ &  &  &  & (2) \\ 
			\hline 
			$\phi^5$ &  &  &  &  \\ 
			\hline 
		\end{tabular} 
		\label{tab:dim5}
	\end{center}
	\vspace{-0.8cm}
\end{table}

\begin{table}[t]
	\begin{center}
		\caption{Application of the non-renormalization theorem to dimension six. The notation is explained in Table I.}
		\vspace{0.3cm}
		\renewcommand{\arraystretch}{1.2}
		\begin{tabular}{|c|c|c|c|c|c|c|c|c|}
			\hline 
			& $F^3$& $\phi^2F^2$ & $F\phi\psi^2$ & $D^2\phi^4$ & $D\phi^2\psi^2$ & $\psi^4$ & $\phi^3\psi^2$ & $\phi^6$ \\ 
			\hline 
			$F^3$ & &\cG $\times_1$ & (2) &\cR $\times_2$ &\cR $\times_2$ &\cR $\times_2$ & \cR $\times_3$ &\cR $\times_3$ \\ 
			\hline 
			$\phi^2F^2$ & &  &  &\cG  &\cG  &  & (2) &\cR $\times_2$ \\ 
			\hline 
			$F\phi\psi^2$ & &  &  &\cG  &\cG  &  &\cG $\times_1$ &\cR $\times_3$ \\ 
			\hline 
			$D^2\phi^4$ &\cG &\cG  &\cG  &  &  &  &\cG $\times_1$ &\cR $\times_2$ \\ 
			\hline 
			$D\phi^2\psi^2$ &\cG &\cG  &\cG  &  &  &  &\cG $\times_1$ & (3) \\ 
			\hline 
			$\psi^4$ & &  &  &  &  &  & (2) & (4) \\ 
			\hline 
			$\phi^3\psi^2$ &  &  &  &  &  &  &  & (2) \\ 
			\hline 
			$\phi^6$ &  &  &  &  &  &  &  & \\ 
			\hline 
		\end{tabular} 
		\label{tab:dim6}
	\end{center}
	\vspace{-0.8cm}
\end{table}

\begin{table}[t]
	\begin{center}
		\renewcommand{\arraystretch}{1.2}
		\caption{Application of the non-renormalization theorem to dimension seven. The notation is explained in Table I. The shortest and longest operators have been dropped from the list of columns and rows, respectively, since our theorem requires a reduction in
			length of the operators. 
		}
		\vspace{0.3cm}
		\begin{tabular}{|c|c|c|c|c|c|c|c|}
			\hline 
			& $\phi^3F^2$ & $D^2\phi^5$ & $D\phi^3\psi^2$ & $\phi\psi^4$ & $F\phi^2\psi^2$ & $\phi^4\psi^2$ & $\phi^7$ \\ 
			\hline 
			$F^3\phi$&\cG $\times_1$ &\cR $\times_2$ &\cR $\times_2$ &\cR $\times_2$ & (2) &\cR $\times_3$ &\cR $\times_3$ \\
			\hline
			$D^2F\phi^3$&\cG $\times_1$ &\cG $\times_1$ &\cG $\times_1$ &\cR $\times_2$ & \cG $\times_1$ &\cR $\times_2$ &\cR $\times_3$ \\  
			\hline 
			$DF\phi\psi^2$& (2) & \cR $\times_2$ &\cG $\times_1$ &\cG $\times_1$ &\cG $\times_1$ &\cR $\times_2$ &\cR $\times_4$ \\ 
			\hline 
			$F^2\psi^2$& (2) & (3) & (2) & (2) &\cG $\times_1$ &\cR $\times_2$ &\cR $\times_4$ \\ 
			\hline 
			$D^2\phi^2\psi^2$& (2) & (2) &\cG $\times_1$ &\cG $\times_1$ &\cG $\times_1$ &\cR $\times_2$ & (4) \\ 
			\hline 
			$D\psi^4$& (3) & (3) & (2) &\cG $\times_1$ & (2) & (3) & (5) \\ 
			\hline 
			$\phi^3F^2$&  & \cG &\cG  &  &  & (2) &\cR $\times_2$ \\ 
			\hline 
			$D^2\phi^5$& \cG &  &  &  & \cG &\cG $\times_1$ &\cR $\times_2$ \\ 
			\hline 
			$D\phi^3\psi^2$& \cG &  &  &  & \cG &\cG $\times_1$ & (3) \\ 
			\hline 
			$\phi\psi^4$&  &  &  &  &  & (2) & (4) \\ 
			\hline 
			$F\phi^2\psi^2$&  & \cG & \cG &  &  &\cG $\times_1$ &\cR $\times_3$ \\ 
			\hline 
			$\phi^4\psi^2$&  &  &  &  &  &  & (2) \\ 
			\hline 
		\end{tabular} 
		\label{tab:dim7}
	\end{center}
	\vspace{-0.75cm}
\end{table}

\Section{Two-loop examples}
Consider now two calculations that show explicit examples, from Table \ref{tab:dim6}, of the nontrivial
zeros in the anomalous-dimension matrix at two loops. The examples will also demonstrate the vanishing of $\gamma_{\rm IR}$. The first example is
the renormalization of $\mathcal O_{\phi^2F^2}$ by $\mathcal
O_{\phi^6}$, which is the entry (2,8) of Table \ref{tab:dim6}. 

The minimal two-loop form factor for $\mathcal O_{\phi^2F^2}$ includes
two external scalars and two external gauge bosons. The product
$\mathcal M F^*$ in \eqn{eq:ffabstract} at two loops requires either a
cut between a five-point amplitude and the tree-level form factor or a
four-point amplitude and a one-loop form factor with an
insertion of $\mathcal O_{\phi^6}$. However, the cut between the five-point
amplitude and the tree-level form factor leaves five total
external legs, and thus cannot match the minimal form factor
for $\mathcal O_{\phi^2F^2}$. For the cut between
the four-point amplitude and the one-loop form factor to match
the minimal form factor for $\mathcal O_{\phi^2F^2}$, the one-loop
form factor would have to involve a massless tadpole, which would
evaluate to zero.

\begin{figure}[tb]
	\centering
	\begin{minipage}{0.49\columnwidth}
	\includegraphics[width=0.9\linewidth]{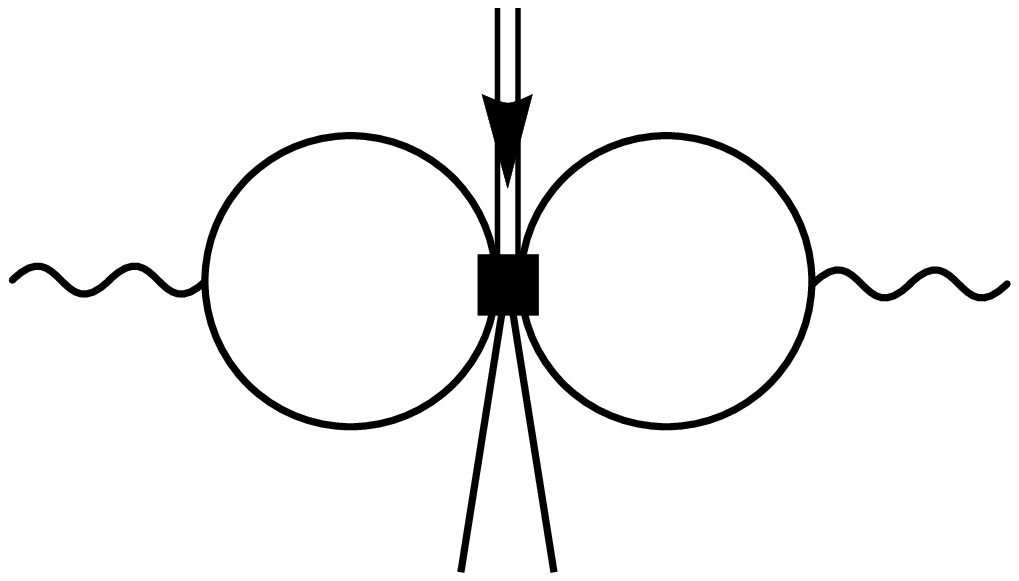}\\[.1 cm] \textbf{(a)}
	\end{minipage}
	\begin{minipage}{0.49\columnwidth}
		\includegraphics[width=0.9\linewidth]{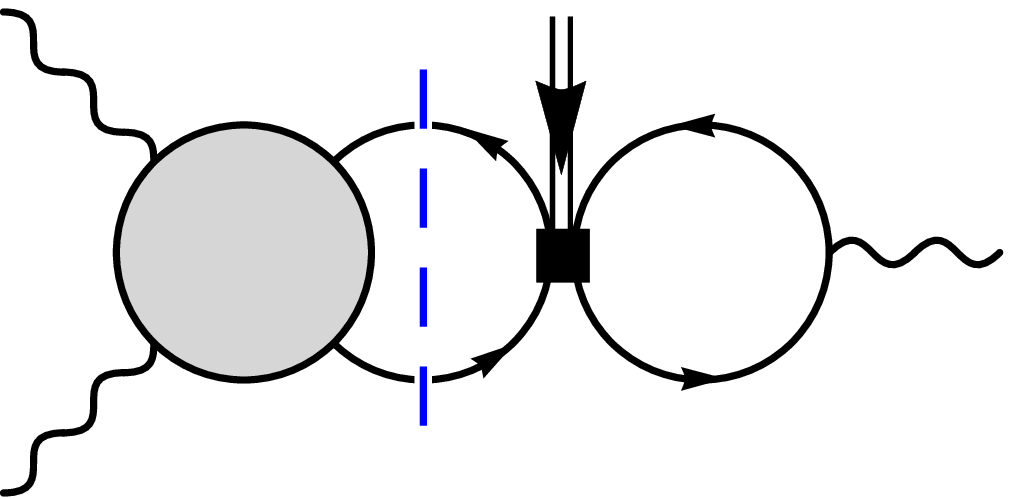}\\[.325 cm] \textbf{(b)}
	\end{minipage}
	\caption{(a) Diagram showing the only possible two loop
          contribution to the renormalization of $\mathcal
          O_{\phi^2F^2}$ by $\mathcal O_{\phi^6}$. (b) Cut of a form factor
          showing that $\mathcal O_{\psi^4}$ cannot renormalize
          $\mathcal O_{F^3}$ at two loops. The solid square indicates the
          insertion of the $\phi^6$ or $\psi^4$ operator,
          respectively.
	} \vspace{-0.5cm}
	\label{fig:formfactorex}
\end{figure}

We can also directly check that the (single)
diagram---\Fig{fig:formfactorex}(a)---for the $\mathcal O_{\phi^6}
\rightarrow \mathcal O_{\phi^2F^2}$ renormalization evaluates to
zero. By incorporating an IR regulator $\lambda_{\rm IR}$, we can evaluate the
integral while keeping the UV and IR dependences separate and
determine the behavior of the form factor in the limit
$\lambda_{\rm IR}\rightarrow0$. The integral for this diagram immediately
factorizes, and each of the two loop integrals is of the form
\begin{equation}
\int\frac{d^D\ell_1}{(2\pi)^D}\frac{(2\ell_1^\nu-k_1^\nu)\, \varepsilon_1^\nu}
{(\ell_1^2-\lambda_{\rm IR})((\ell_1-k_1)^2-\lambda_{\rm IR})}\,.
\end{equation}
This integral vanishes by the on-shell condition $k_1
\cdot\varepsilon_1=0$ and Lorentz invariance, since $k_1$ is the only
available momentum. Therefore $\mathcal O_{\phi^6}$ cannot renormalize
$\mathcal O_{\phi^2F^2}$ at two loops.

For a slightly more complex example, consider the renormalization of
$\mathcal O_{F^3}$ by $\mathcal O_{\psi^4}$ at two loops, corresponding to entry (1,6) of Table \ref{tab:dim6}. Again, for
this process the three-particle cut between the five-point amplitude and the tree-level
form factor does not produce the correct external-particle state
corresponding to the field content of $\mathcal O_{F^3}$. The
two-particle cut between the four-point amplitude and the one-loop form factor with an
insertion of $\mathcal O_{\psi^4}$ is shown in Figure
\ref{fig:formfactorex}(b). By again adding an IR regulator, the result
can be written as
\begin{equation}
  \int d\text{LIPS}_{\ell_1} \frac{ d^D\ell_2}{(2\pi)^D}\frac{\Tr[X(\ell_1)\slashed \ell_2\slashed \varepsilon_3 (\slashed \ell_2-\slashed k_3)]}{(\ell_2^2-\lambda_{\rm IR})((\ell_2-k_3)^2-\lambda_{\rm IR})} \,,
  \label{eq:examplebubble}
\end{equation}
where $X$ receives contributions from the multiple possible diagrams of the four-point amplitude and includes the remaining propagators. LIPS indicates that the integration is over the Lorentz-invariant phase space of the particles crossing the cuts.

One can reduce the $\ell_2$ tensor integrals using standard techniques to obtain the following result 
  \vspace{-0.1cm}
\begin{align}
  \int\frac{d^D\ell_2}{(2\pi)^D}&\frac{\ell_2^\mu\ell_2^\nu}{(\ell_2^2-\lambda_{\rm IR})^2}\int d\text{LIPS}_{\ell_1}\, Y_{\mu\nu}(\ell_1) \\
  =&-\frac{i\Gamma(-1+\epsilon)}{2(4\pi)^{2-\epsilon}}(\lambda_{\rm IR})^{1-\epsilon}\int d\text{LIPS}_{\ell_1}\, Y^{\mu}{}_{\mu}(\ell_1) \,, \nonumber \vspace{-0.2cm} 
\end{align}
where $\epsilon=(4-D)/2$, $Y$ contains the rest of the trace in
Eq.~\eqref{eq:examplebubble}, and terms linear in $\ell_2$ cancel.
Since the phase-space integral can at worst result in a
$\log(\lambda_{\rm IR})$ divergence, the factor $(\lambda_{\rm
  IR})^{1-\epsilon}$ ensures that the expression goes smoothly to zero
as $\lambda_{\rm IR}$ approaches zero for all orders in
$\epsilon$. Therefore the cut vanishes, along with the UV anomalous dimension.

\Section{Conclusions}
We have derived a new non-renormalization theorem that applies to
higher-dimensional operators in quantum field theory. Since
the theorem is dependent on only the number and type of fields in each
operator, it applies to generic massless theories with no relevant
operators.

Besides being helpful to find zeros of the anomalous-dimension matrix,
the on-shell formalism of Ref.~\cite{Caron-Huot:2016cwu} is a good way
to compute nonzero entries as well.  Whenever an entry is excluded by our theorem, it should be
much simpler to compute the entry at the next loop order compared to
computing a generic entry at that loop order, because only tree-level
quantities enter the cuts.  In addition, helicity
selection rules~\cite{NonrenormHelicity} might then apply, 
pushing the zero one loop further. For instance, it is
straightforward to confirm that many of the nonzero entries in the
tables above vanish in the absence of Yukawa couplings.
It would also be interesting to combine
our results with those of Ref.~\cite{AneeshBounds}, where
dimensional-analysis counting rules are used to
constrain coupling-constant dependence, and more generally to find the full set constraints in the multiloop
anomalous-dimension matrix of the SMEFT.  On-shell
methods~\cite{GeneralizeUnitarity} are also a good way to compute
amplitudes including higher-dimension operators.  Using these we have
computed four-point one-loop massless
amplitudes and associated anomalous dimensions of the SMEFT
dimension-six operators, which will be described
elsewhere~\cite{OneLoopSMEFT}.

\Section{Acknowledgments}
We thank Clifford Cheung, Enrico Herrmann, Aneesh Manohar, Ian Moult,
Chia-Hsien Shen and George Sterman for very helpful discussions.  We
thanks the U.S. Department of Energy (DOE) for support under grant
no.~DE-SC0009937. We are also grateful to the Mani L. Bhaumik
Institute for Theoretical Physics for generous support.

\end{document}